\documentclass[review]{elsarticle}

\usepackage{lineno,hyperref}
\modulolinenumbers[5]

\usepackage{upgreek}
\usepackage{subfigure}
\usepackage{textcomp}
\usepackage{amssymb}
\usepackage{amsthm}

\usepackage{bm}

\newcommand{\uvec}[1]{\boldsymbol{\hat{\textbf{#1}}}}
\usepackage{esint}

\begin{document}
\begin{frontmatter}

\title{Convolution based hybrid image processing technique for microscopic images of etch-pits in Nuclear Track Detectors}

\author{Kanik Palodhi}
\author{Joydeep Chatterjee}
\address{Department of Applied Optics and Photonics, University of Calcutta, Kolkata 700106, India}
\author{Rupamoy Bhattacharyya\footnote{Corresponding author. \\E-mail address: rupamoy@gmail.com\\ Present address: Indian Institute of Science Education and Research Bhopal, Madhya Pradesh 462066, India}}
\author{S. Dey}
\author{Sanjay K. Ghosh}
\author{Atanu Maulik\footnote{Present address: Istituto Nazionale di Fisica Nucleare, Sezione di Bologna, Bologna 40127, Italy}}
\author{Sibaji Raha}
\address{Centre for Astroparticle Physics and Space Science, Bose Institute, Kolkata 700 091, India}

\begin{abstract}
A novel image processing technique based on convolution is developed for analyzing the etch-pit images in Nuclear Track Detectors (NTDs). The outcomes of the application of the proposed method on the different types of NTDs (e.g., CR-39, PET) containing etch-pit openings of different sizes and shapes (circular and elliptical) is presented. Promising results have been obtained for both identifying and counting the etch-pits in NTDs.

\end{abstract}

\end{frontmatter}

\section{Introduction}
Nuclear Track Detectors (NTDs) have been used in charged particle detection for many decades in fields ranging from physics to geology~\citep{Fleischer:1975ya,CECCHINI2008S144}. NTDs are dielectric solids (e.g., polymer sheets of thickness $\sim100~\upmu$m). A charged particle, while passing through the NTD material, loses energy by ionizing the medium. If the energy loss is above a certain threshold, then the particle leaves behind a permanent damage trail, essentially broken polymer chains in case of plastics, called a \textquotedblleft latent track\textquotedblright. Obviously, this threshold will be different for different NTD materials. Such damaged regions become chemically more reactive compared to the undamaged bulk material. When such an NTD containing any \textit{latent track} is treated with a suitable chemical reagent, called etchant, materials along the damage trail are etched out at a much faster rate compared to the surrounding bulk material, resulting in etch pits large enough (of the order of micron) to be observed under optical microscopes. The geometry of such etch-pits can reveal crucial information on the identity of the particles forming such tracks. Because of their low cost, ease of handling and existence of natural thresholds of registration (which helps in reducing the background), NTDs are often the detectors of choice in the search for rare heavily ionizing hypothesized particles (e.g., magnetic monopoles, strangelets) in cosmic rays as well as particle accelerators~\citep{Acharya2016,Balestra:2008ps}. It may be mentioned here we are aiming to use a low cost, commercially available polymer, identified as Polyethylene Terephthalate (PET), as NTD in the search for strangelets through the deployment of large-area arrays at mountain altitudes~\citep{1475-7516-2017-04-035}. 
\par
In such searches employing NTDs, the task of scanning of etched NTDs to locate the etch-pit openings on their surface is extremely labour intensive as the researchers are required to scan large areas of NTDs under high magnification. 
The difficulty of finding a track due to any rare event is compounded by the presence of background, which can come from other ionizing radiations and also from structural defects in the plastic which creeps in during the polymerization process. Therefore, challenges of rare event search with NTDs are primarily technological with the conventional image analysis software often coming up short in the task of track identification. 
\par
In this paper, we are proposing a novel approach for etch-pit image identification and counting in NTDs, which shows much-improved performance compared to more \textquotedblleft classical\textquotedblright~ (employing cuts based on just grey level, size, etc.) approaches to image analysis.

\section{Experimental technique}
\begin{figure}[h]
\centering
\includegraphics[width=250px,height=200px]{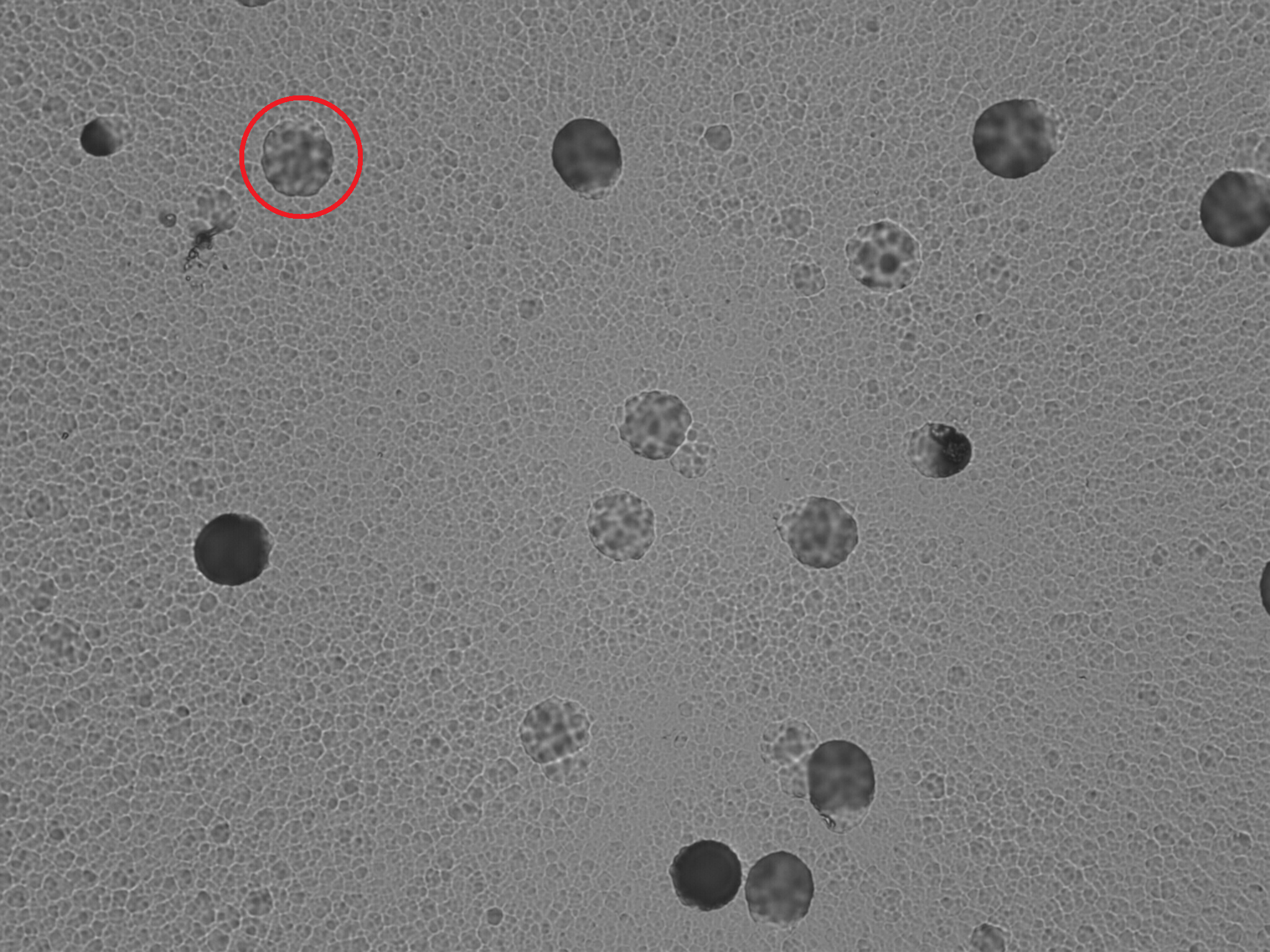}
\includegraphics[width=250px,height=200px]{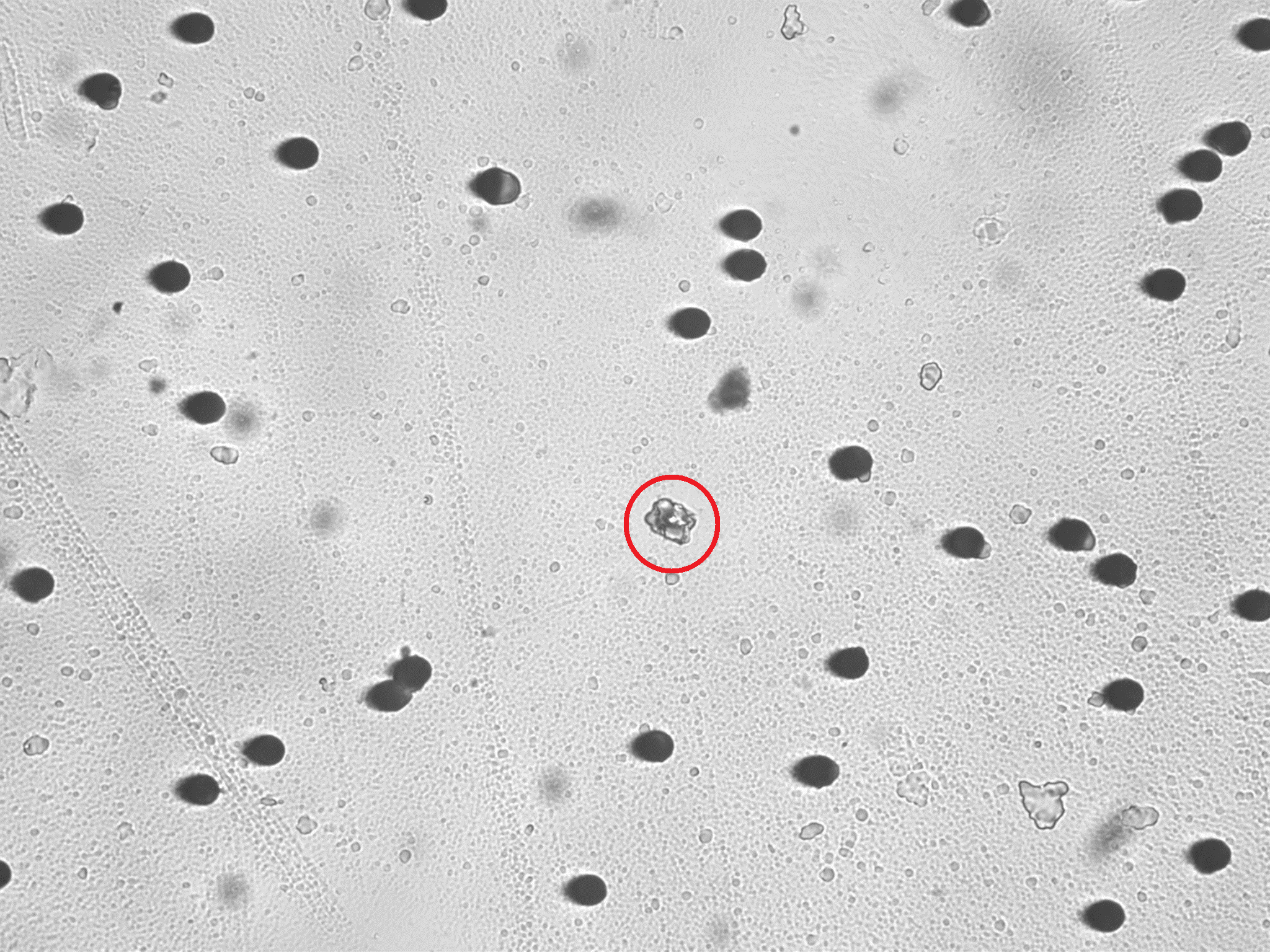}
\caption{Microscopic surface images at 50x objective magnification of (a) CR-39 (NTD) exposed to open-air at Darjeeling after 4 h of etching, where the etch-pits are formed due to local radon alphas, neutron recoiling, cosmic rays and (b) PET (NTD) exposed to 3.7 MeV/nucleon scattered $~^{32}$S ions at an angle $30^{\circ}$ from accelerator after 3 h of etching. The sizes of the image frames are $234~\upmu$m $\times$ $174~\upmu$m. Typical scratches and/or other defects are shown inside the red circle. It may be noted here that the surface quality of the NTDs which get open-air exposure are worsened due to harsh environmental conditions.}
\label{fig1} 
\end{figure}

This section describes the image analysis techniques that are applied to the unprocessed surface images of exposed and etched NTDs, captured by QWin software using a Leica DM4000B optical microscope. The etch-pits appear dark as compared to the background of NTD surface; however, the challenge is to separately recognize them from the other artifacts generated during chemical etching and scratches and defects, which also appear dark, as shown in Fig~\ref{fig1}. For this study, we have used the etch-pit images from Columbia Resin \#39 (CR-39) and Polyethylene terephthalate (PET) NTDs exposed to open-air at Darjeeling in Eastern Himalayas (altitude 2.2 km above mean sea level), India~\citep{1475-7516-2017-04-035} and $3.9$ MeV/nucleon $~^{32}$S ions from pelletron accelerator at Inter University Accelerator Centre, New Delhi~\citep{BHOWMIK2011197}, respectively. It may be mentioned here that for accelerator exposed NTDs (e.g., PET in this case), the bulk etch-rate ($V_B$) remains the same with that of the unexposed one (i.e., for PET $V_B=1.0\pm0.05~\upmu$m/h at $55.0\pm0.1$ $^{\circ}$C in 6.25 N NaOH aqueous solution), whereas typical values of $V_B$ for CR-39 increased by several factors (from $V_B=1.4\pm0.07~\upmu$m/h for unexposed one to $V_B=11.7\pm0.7~\upmu$m/h at $70.0\pm0.1$ $^{\circ}$C in 6.25 N NaOH aqueous solution) as the surface undergoes some degradation because of prolonged open-air exposure. 

\par
The image analysis technique that we have applied here is based on sequential application of blind de-convolution and convolution with a suitable mask size obtained from the analysis of the area occupied by the two-dimensional elliptical opening (circular opening in case of normally incident ion) of the three-dimensional etch-pit cone. Before going to the application of the technique, a few considerations regarding image quality and image acquisition are described below:
\par
The typical sizes of etch-pit openings used here are roughly $\sim1-10~\upmu$m. As mentioned earlier, there are artifacts of similar sizes, which may be wrongly counted. Convolutions with the original size of the opening of etch-pits, therefore, need to be augmented with the shape of the pit-openings. It is known that the typical shape of the opening of the etch-pits will be elliptical, and in special cases circular (for normal incidence of the incoming ion), as shown in Fig.\ref{fig5} and Fig.\ref{fig4} respectively.
\par

\begin{figure}[h]
\centering
\includegraphics[width=180px,height=120px]{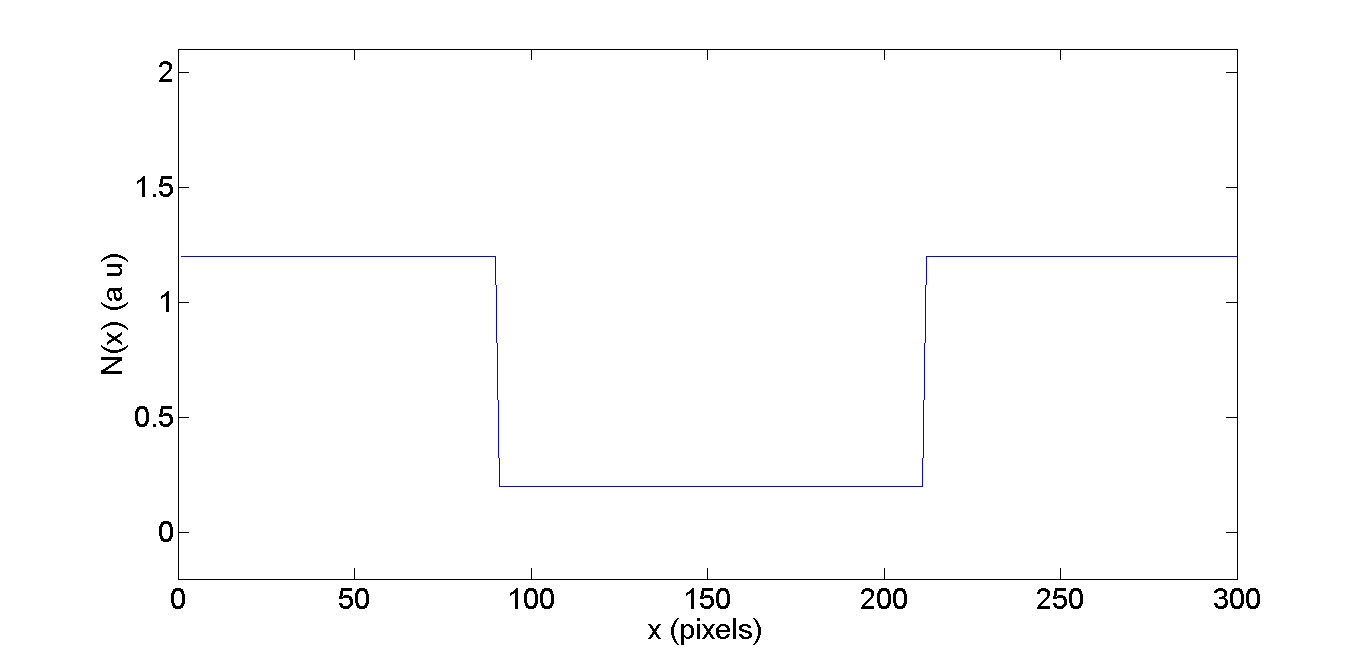}
\includegraphics[width=180px,height=120px]{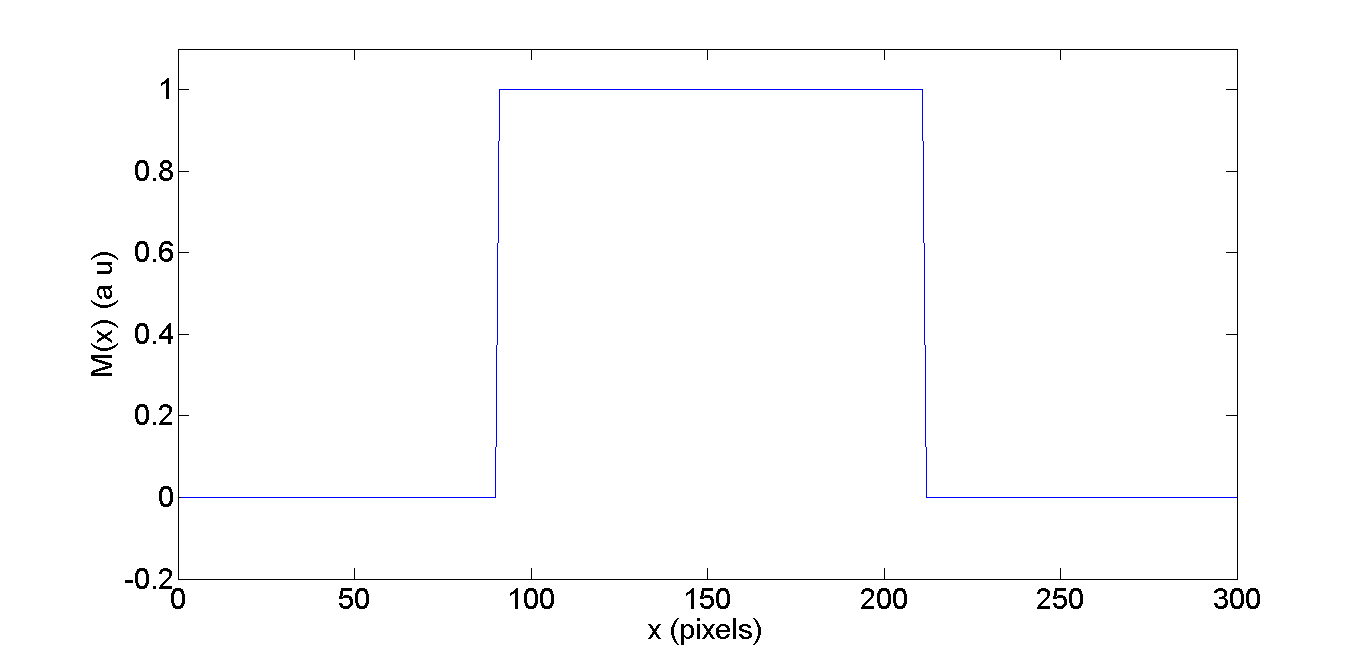}
\includegraphics[width=180px,height=120px]{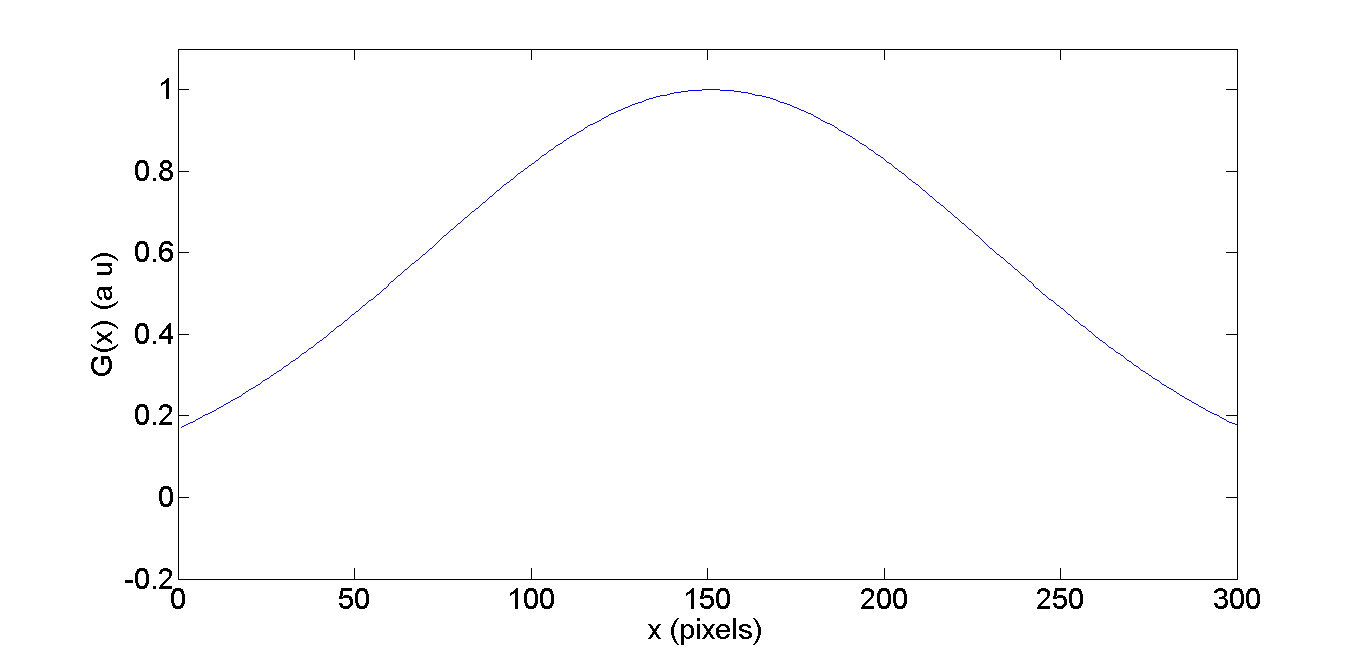}
\includegraphics[width=180px,height=120px]{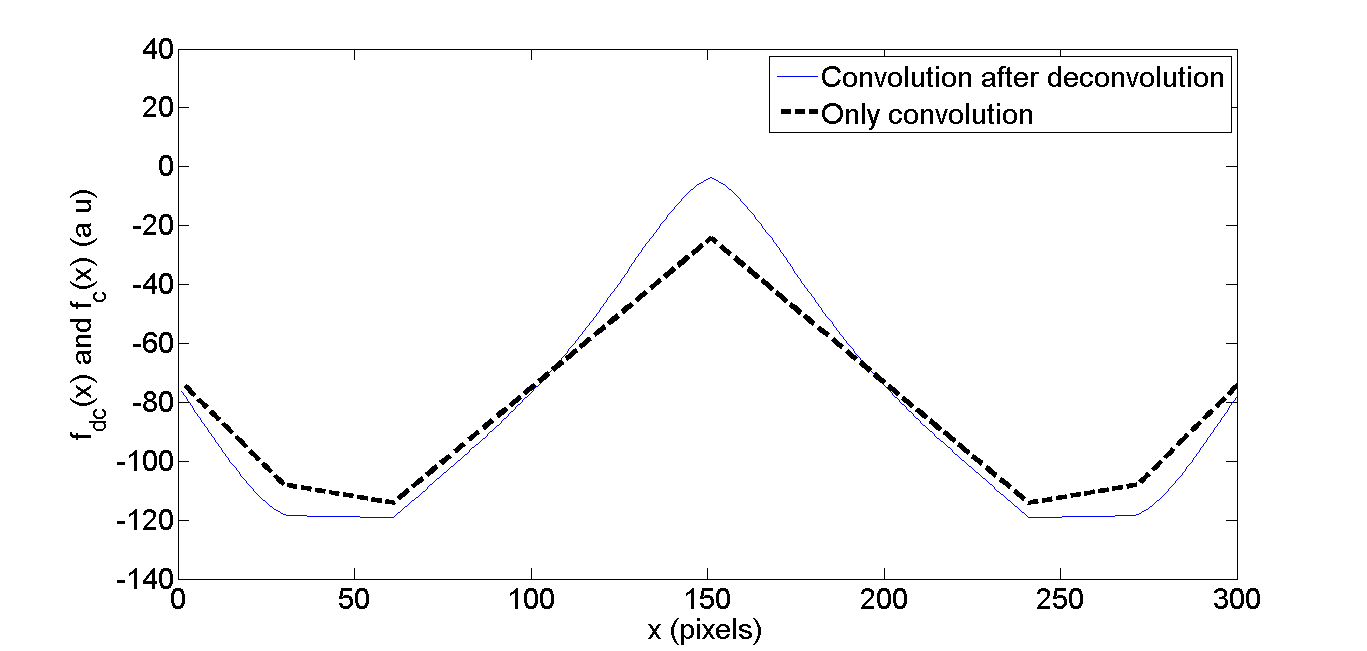}
\caption{One dimensional representation of only convolution and then de-convolution followed by convolution process. (a) One dimensional representation of the opening of etch-pit profile $N(x)$ (width $\sim120$ pixel); (b) The circular mask $M(x)$ (width $\sim120$ pixels); (c) The Gaussian mask used in de-convolution $G(x)$ (half-width $\sim120$ pixels); (d) Comparison of convolution $f_c(x)$ and de-convolution followed by convolution $f_{dc}(x)$.}
\label{fig2} 
\end{figure}

First of all, among all the present etch-pit openings, the biggest one in shape and size is ascertained. Then a similar convolution circular mask is formed which is convolved with the entire etch-pit opening. As discussed, it provides a huge advantage since the convolution peak, generated at the centre of the etch-pit, helps to determine the position of a track. If any of the etch-pit openings within the captured image can be expressed as $N(x,y)$, considering its two-dimensional form and the circular mask mentioned above, is expressed as $M(x,y)$, then their convolution can be written as
\begin{equation}
f_c (x,y) = \iint \limits_{-\infty}^{+\infty}  N(x-x_0,y-y_0)M(x_0,y_0)dx_0dy_0
\end{equation}
The above expression is commonly written as $f_c(x,y)=N(x,y) * M(x,y)$ and is routinely used in different domains of optics and signal processing~\citep{Gaskill_1978}. The advantage of using convolution is that in the case of the circular mask convolving with circular or elliptical opening of etch-pits, the peak values are produced at the centre provided that the shapes and sizes are nearly equal. In general, this is true for NTD surfaces. As a typical example, the average diameter of the etch-pit openings shown in Fig.\ref{fig1}(a) corresponds to nearly 120 pixels ($\sim12~\upmu$m). Fig.\ref{fig2} is an illustration by simulation where a one-dimensional representation of convolution is shown between an etch-pit and a mask of the same size. In one-dimensional representation, the functions are represented with a single variable provided in terms of pixels with an analogy to the images. Therefore, when $N(x)$ (Fig.\ref{fig2}(a)), essentially a one-dimensional projection of the etch-pit opening profile, gets convolved with the mask $M(x)$ (Fig.\ref{fig2}(b)), the result is a well-known triangular function as shown in Fig.\ref{fig2}(d) with a dotted line.
\begin{equation}
f_c (x,y) = \iint \limits_{-\infty}^{+\infty}  N(x-x_0,y-y_0)M(x_0,y_0)dx_0dy_0
\end{equation}
The above expression is commonly written as $f_c(x,y)=N(x,y) * M(x,y)$ and is routinely used in different domains of optics and signal processing~\citep{Gaskill_1978}. The advantage of using convolution is that in the case of the circular mask convolving with circular or elliptical opening of etch-pits, the peak values are produced at the centre provided that the shapes and sizes are nearly equal. In general, this is true for NTD surfaces. As a typical example, the average diameter of the etch-pit openings shown in Fig.\ref{fig1}(a) corresponds to nearly 120 pixels ($\sim12~\upmu$m). Fig.\ref{fig2} is an illustration by simulation where a one-dimensional representation of convolution is shown between an etch-pit and a mask of the same size. In one-dimensional representation, the functions are represented with a single variable provided in terms of pixels with an analogy to the images. Therefore, when $N(x)$ (Fig.\ref{fig2}(a)), essentially a one-dimensional projection of the etch-pit opening profile, gets convolved with the mask $M(x)$ (Fig.\ref{fig2}(b)), the result is a well-known triangular function as shown in Fig.\ref{fig2}(d) with a dotted line.

\par

\begin{figure}[h]
\centering
\includegraphics[width=150px,height=150px]{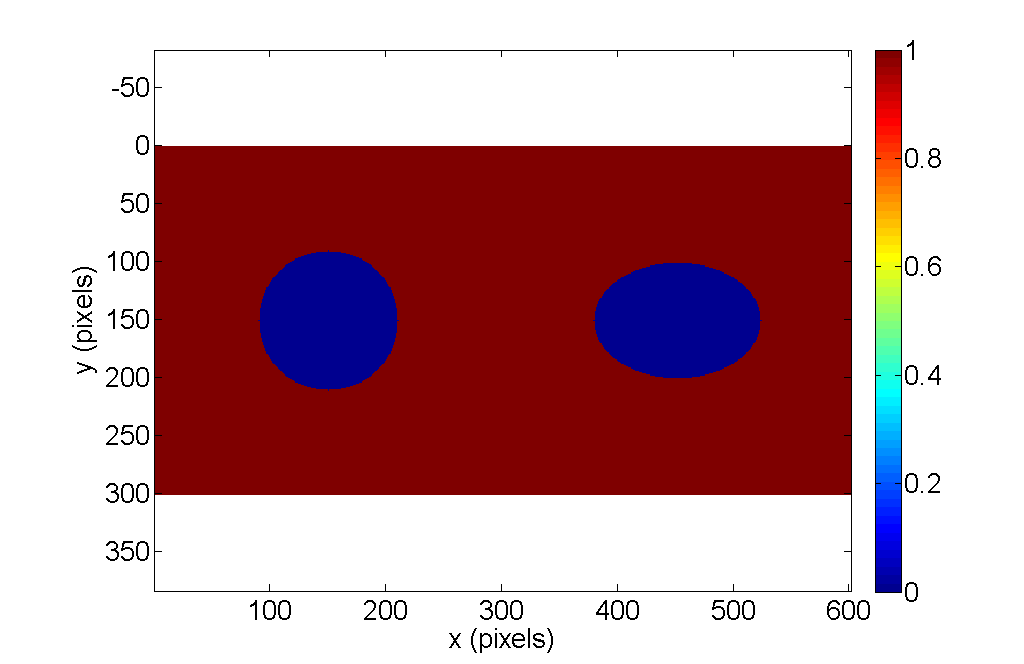}
\includegraphics[width=150px,height=150px]{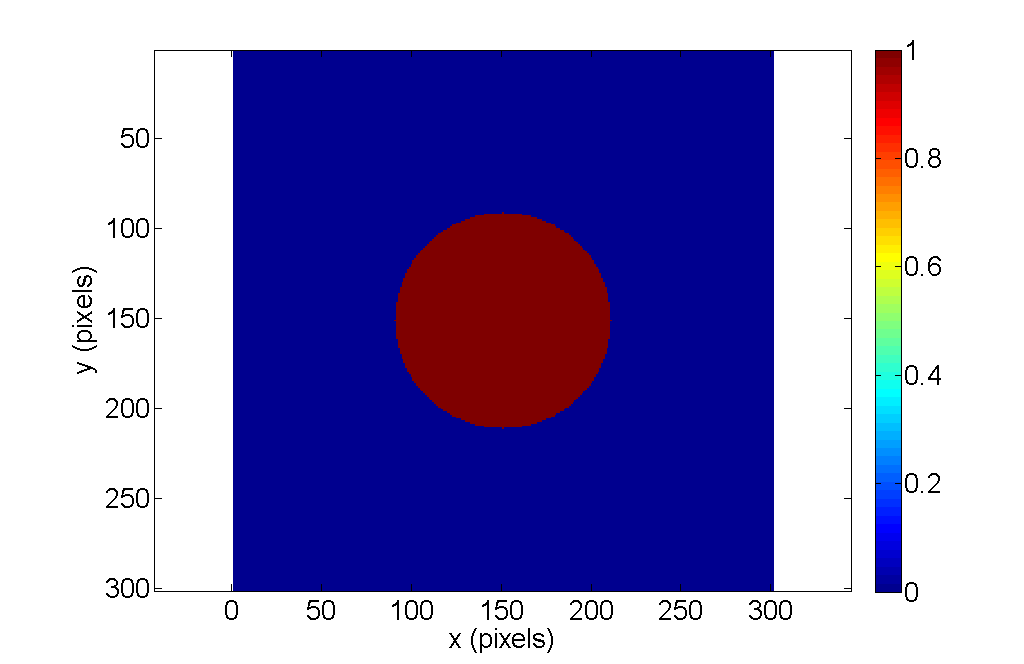}
\includegraphics[width=180px,height=150px]{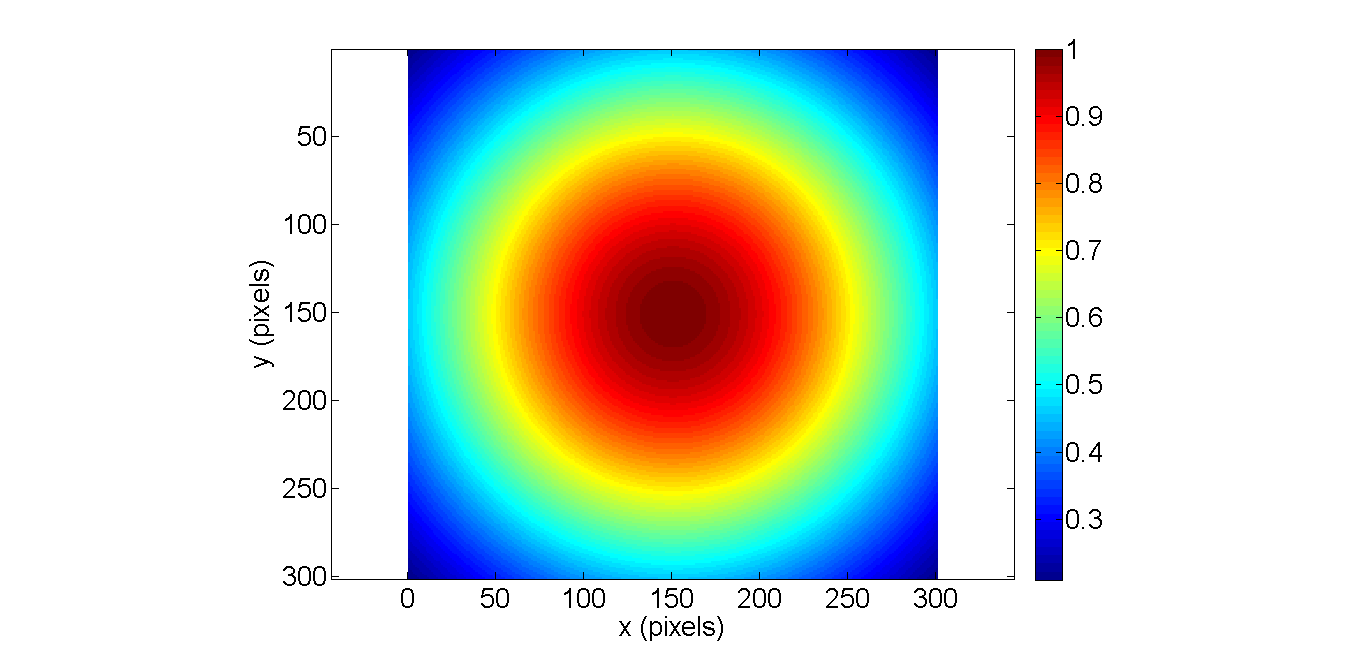}
\includegraphics[width=150px,height=150px]{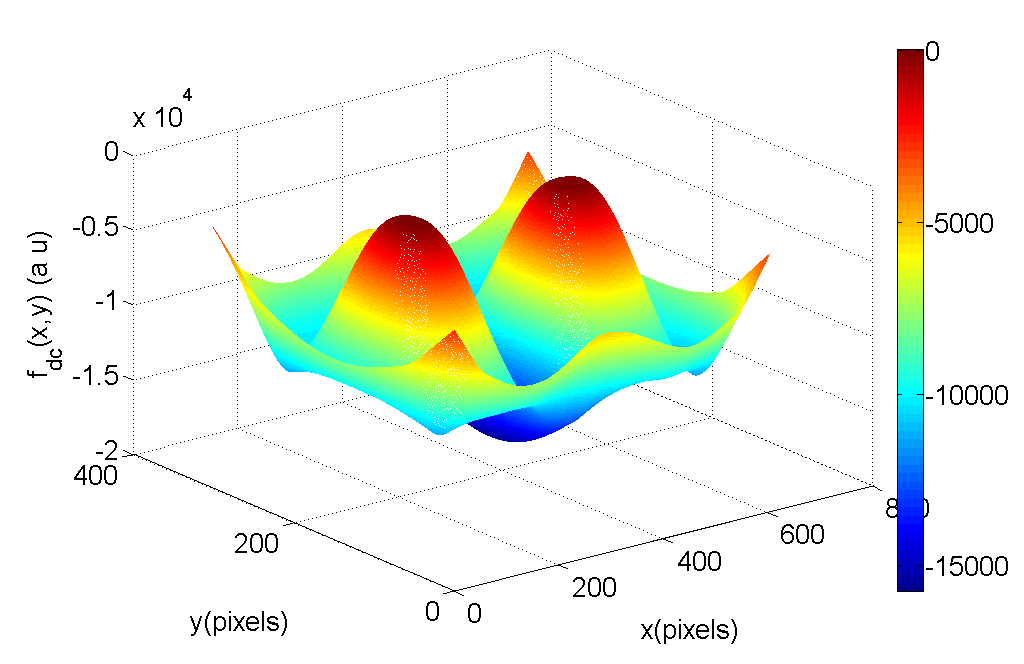}
\includegraphics[width=150px,height=150px]{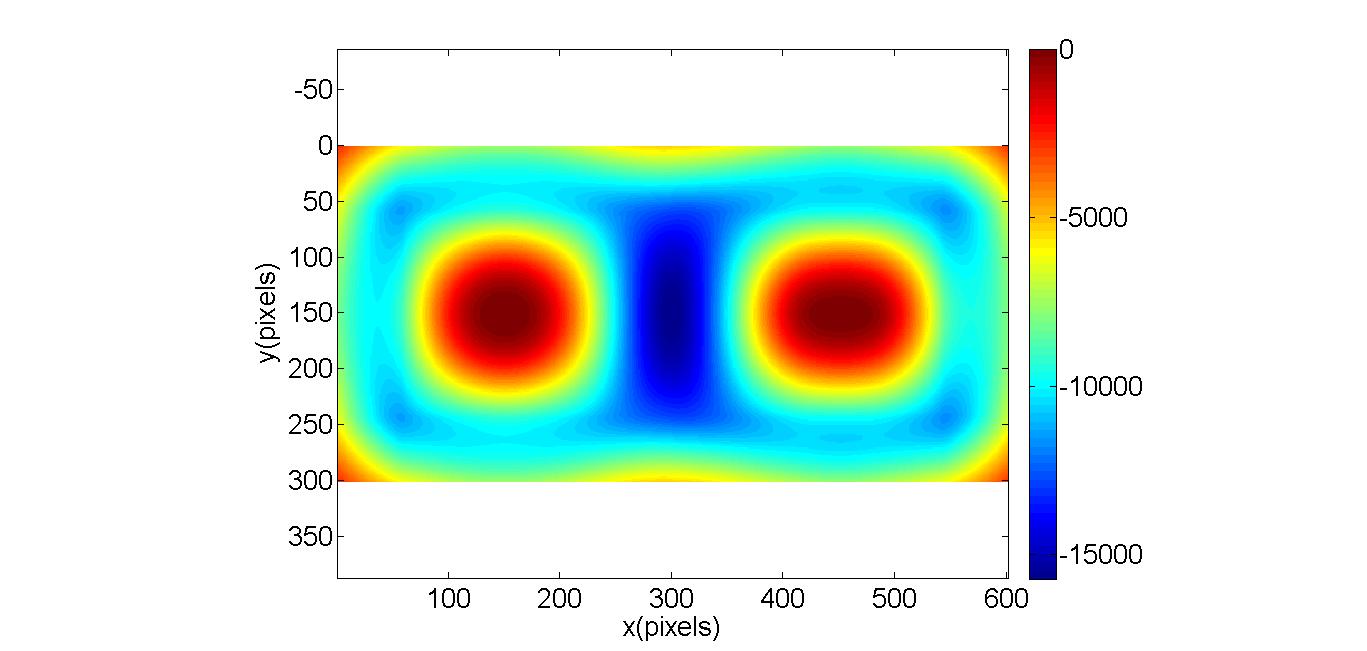}
\caption{Two dimensional representation of only convolution and then de-convolution followed by convolution process. (a) Simulated images of openings of etch-pits of both circular and elliptical shape (diameter $\sim120$ pixel); (b) The circular mask (diameter $\sim120$
pixels); (c) The Gaussian mask used in de-convolution (half-width $\sim120$ pixels); (d) De-convolution followed by convolution $f_{dc}(x)$. (e) Two-dimensional projection of $f_{dc}(x)$ for circular and elliptical etch-pit opening.}
\label{fig3} 
\end{figure}

For actual cases in NTDs, however, the other artifacts or the scratches are of different shapes and sizes, will result in similar convolution peaks. Therefore, for enhancing the peak values near the centres of etch-pit openings, a de-convolution process with a Gaussian mask as shown in Fig.\ref{fig2}(c) of similar shape (half-width $\sim 120$ pixels) is introduced~\citep{Ulmer_2013}. This provides higher peaks compared to simple convolution for definitive shapes of NTDs as shown in Fig.\ref{fig2}(d). Theoretically, this can be represented considering a Fourier domain explanation where $G(x)$ (Fig.\ref{fig2}(c)) is a one-dimensional representative Gaussian function centred at origin given by
\begin{equation}
G(x)=\exp(-ax^2)
\end{equation}
The Fourier transform of the above Gaussian function, is also a Gaussian function when written in terms of spatial frequency along X-direction $k_x$
\begin{equation}
F\{G(x)\}= \uvec{G}(k_x)=\sqrt{\frac{\pi}{a}}\exp(\frac{-\pi^2{k_{x}^{2}}}{a})
\end{equation}
Therefore, de-convolution in Fourier domain can be expressed as
\begin{equation}
\label{eqn4}
\uvec{f}_d(k_x)=\frac{\uvec{N}(k_x)}{\uvec{G}(k_x)}
\end{equation}
where $\uvec{f}_d(k_x)$ and $\uvec{N}(k_x)$ represent the Fourier transforms of the respective functions introduced already. As described in Ref.~\citep{Ulmer_2013}, the technique of applying Gaussian in de-convolution reduces a lot of problems since the characters of the functions in both domains are well-known~\citep{Ulmer_2013}. Next, inverse Fourier transform is applied followed by another convolution operation with the standard circular mask described previously (in its one dimensional representation) to enhance the peak at the centre of the opening of etch-pit as shown in Fig.\ref{fig2}(d). This is expressed in terms of convolution operator as
\begin{equation}
\label{eqn6}
f_{dc}(x)=f_d(x) * M(x)
\end{equation}
where$f_d(x)$ is the is inverse Fourier transform of $\uvec{f}_d(k_x)$. A comparison is shown in Fig.\ref{fig3}(d) which clearly indicates that $f_{dc}(x)$, function due to Gaussian de-convolution followed by convolution (Eq.~(\ref{eqn6})) with mask, produces higher peaks compared to simple convolution operation represented by $f_c(x)$. Without any loss of generality, the two-dimensional simulation results corresponding to the above operations are presented in Fig.\ref{fig3}, where the effects of shape variations of etch-pit openings are also considered.

In the case of actual microscopic NTD surface images, the background being whitish, the dark openings of etch-pits have a significantly higher signal to noise ratio where this operation produces a better result. Another advantage is that in cases of noise removal of corresponding frequency domain processing can be done since the convolution process lends itself easily to convolution and de-convolution.

\section{Image analysis and results}
\begin{figure}[h]
\centering
\includegraphics[width=400px,height=200px]{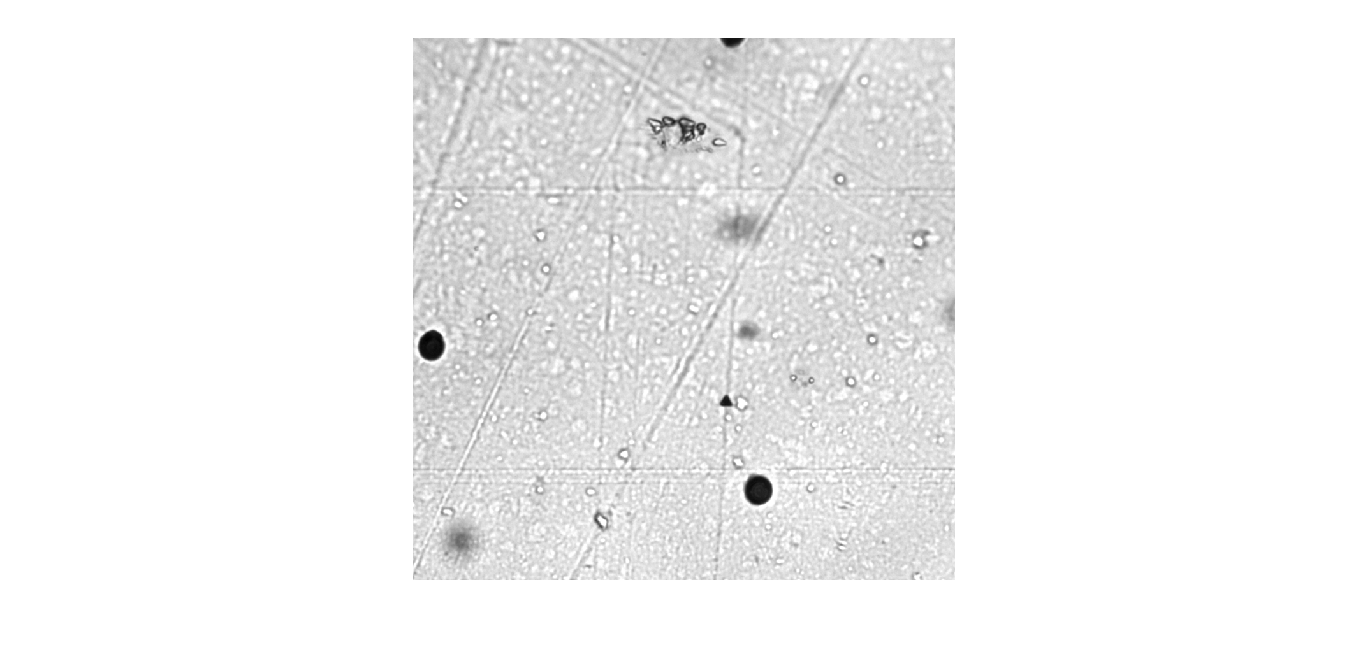}
\includegraphics[width=400px,height=200px]{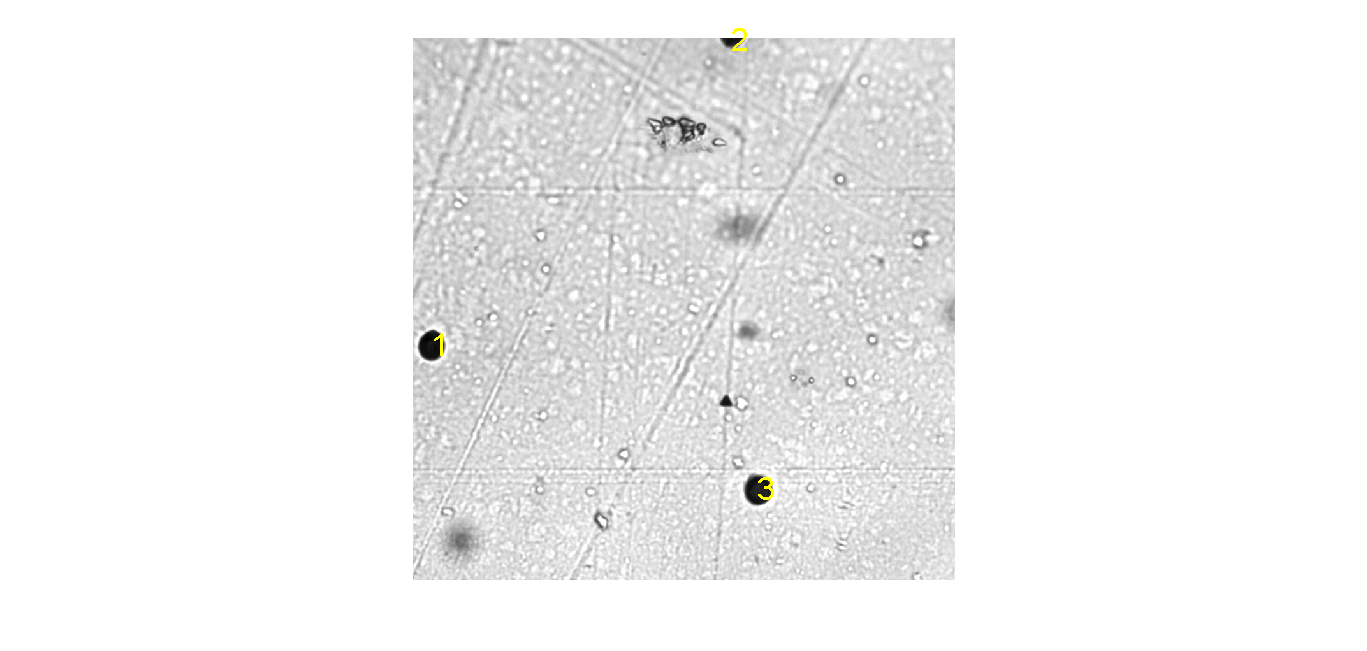}
\caption{(a) Etch-pit openings due to normally incident 3.4 MeV/nucleon scattered $~^{32}$S ions on PET after 2 h of etching at the above mentioned etching conditions. This image also shows scratches (dark as compared to the whitish background); (b) Judicially counting of etch-pits among multiple types of defects. The etch-pits are correctly identified from the background. The size of the image frames are $97~\upmu$m $\times$ $97~\upmu$m.}
\label{fig4} 
\end{figure}

\begin{figure}[h]
\centering
\includegraphics[width=400px,height=200px]{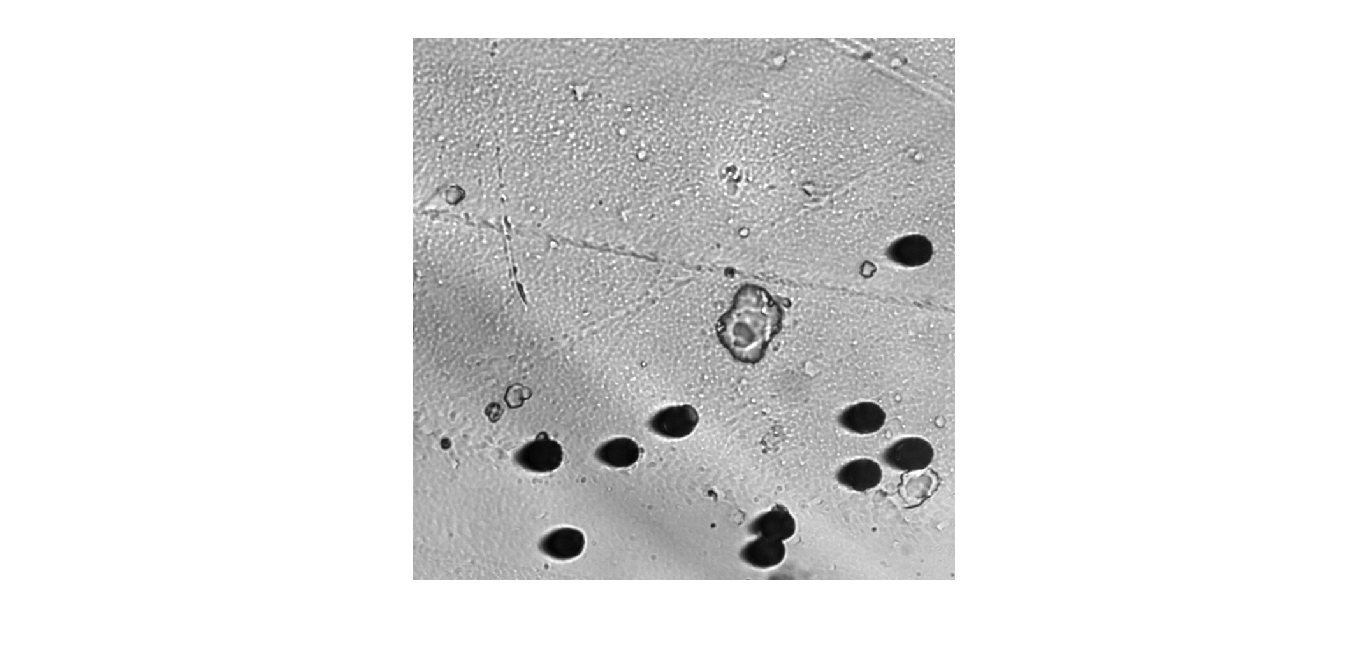}
\includegraphics[width=400px,height=200px]{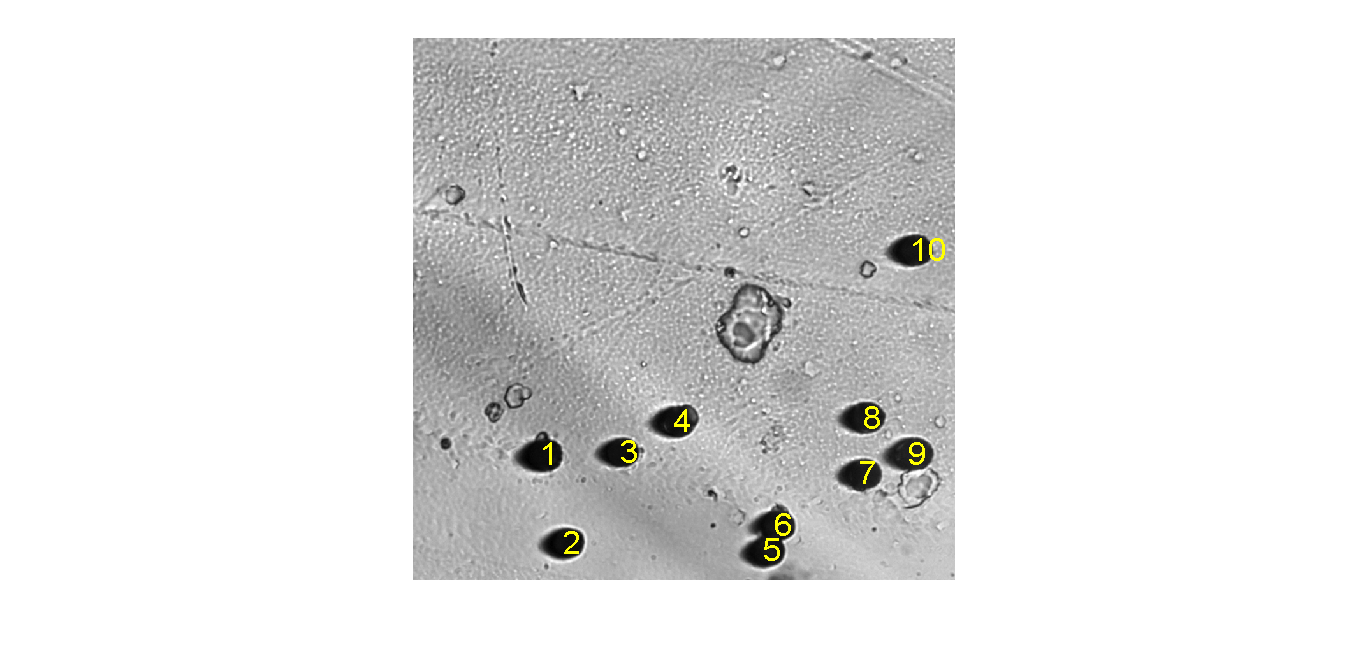}
\caption{(a) Etch-pit openings on PET after 3 h of etching due to the 3.7 MeV/nucleon scattered $~^{32}$S ions striking at $30^{\circ}$ angle of incidence, showing darker etch-pit openings and defects or scratches; (b) Counting of etch-pits separately from multiple types of defects. The size of the image frames are $97~\upmu$m $\times$ $97~\upmu$m.}
\label{fig5} 
\end{figure}

The novel image analysis technique employed here is tested with multiple
microscopic images of NTDs using the copyrighted software Cell Counter~\citep{name}. As described before, images of etch-pit openings due to accelerator exposed NTDs (Fig.\ref{fig4} and Fig.\ref{fig5}) are often used as reliable guidance. Fig.\ref{fig4}(a) shows an image frame with a normal incidence of ions from accelerator; the image has multiple defects and scratches, even of sizes similar to that of the actual etch-pit openings. Clearly, the etch-pits have been separated from the rest and counted (as well as marked) correctly as shown in Fig.\ref{fig4}(b).

\begin{figure}[h]
\centering
\includegraphics[width=400px,height=200px]{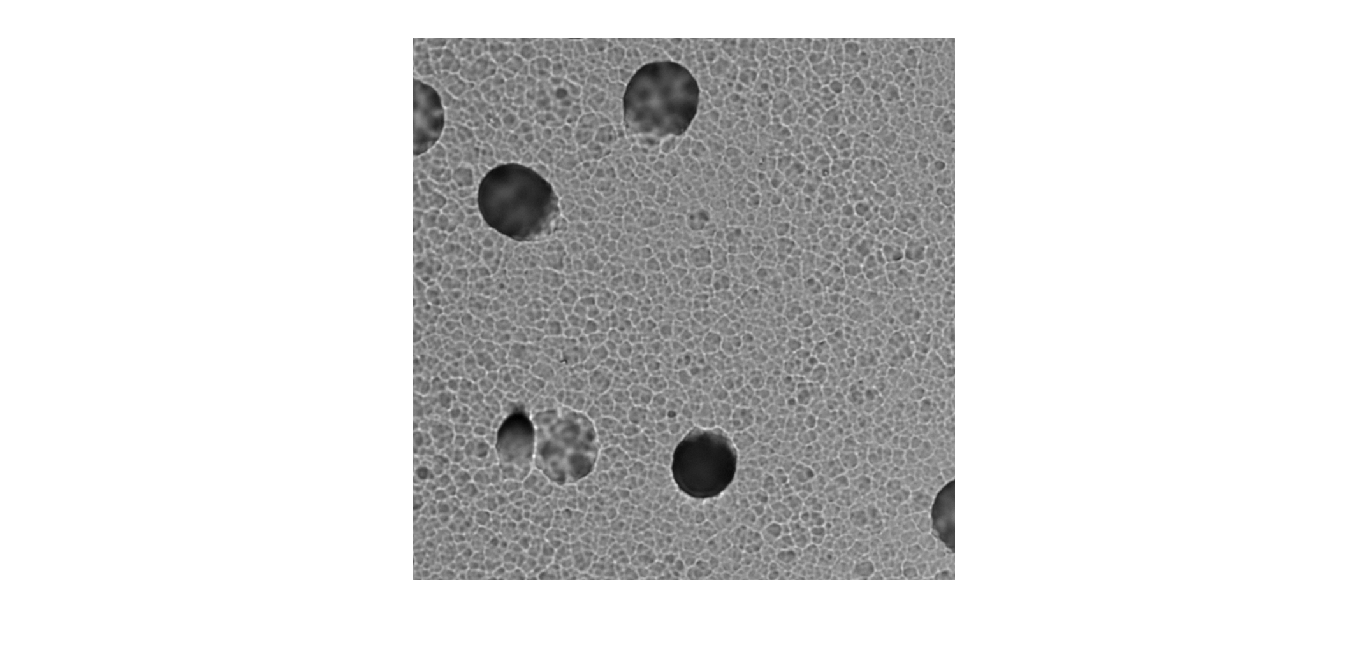}
\includegraphics[width=400px,height=200px]{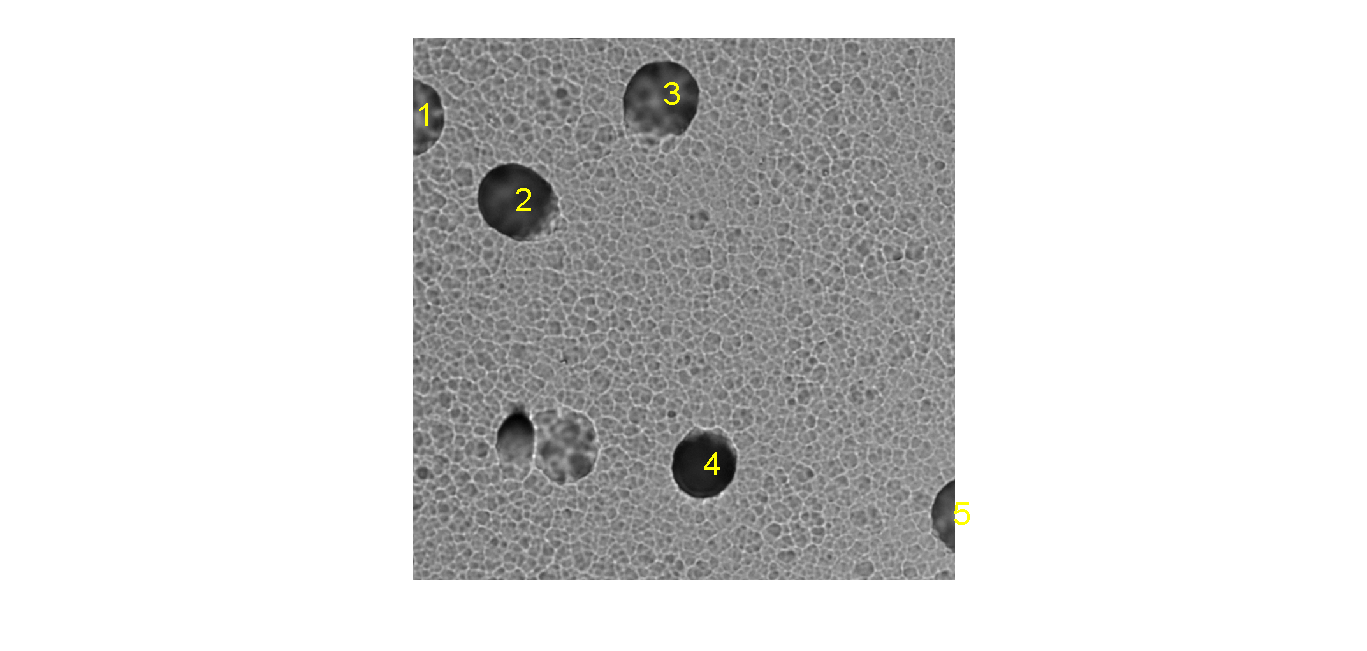}
\caption{(a) Etch-pit openings due to charged particles with different angle of incidences on CR-39 (NTD) exposed to open-air at Darjeeling, India after 4 h of etching; (b) Counting of etch-pits separately from multiple-type of defects. The size of the image frames are $97~\upmu$m $\times$ $97~\upmu$m.}
\label{fig6} 
\end{figure}

\begin{figure}[h]
\centering
\includegraphics[width=400px,height=200px]{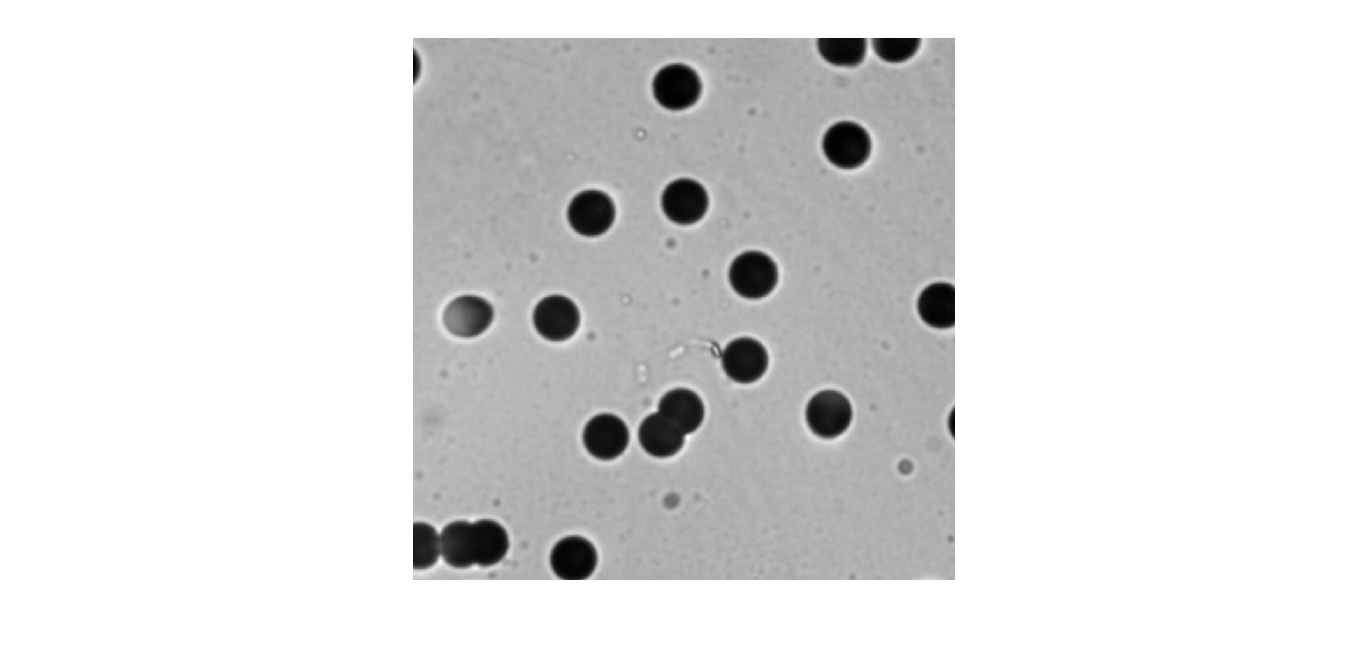}
\includegraphics[width=400px,height=200px]{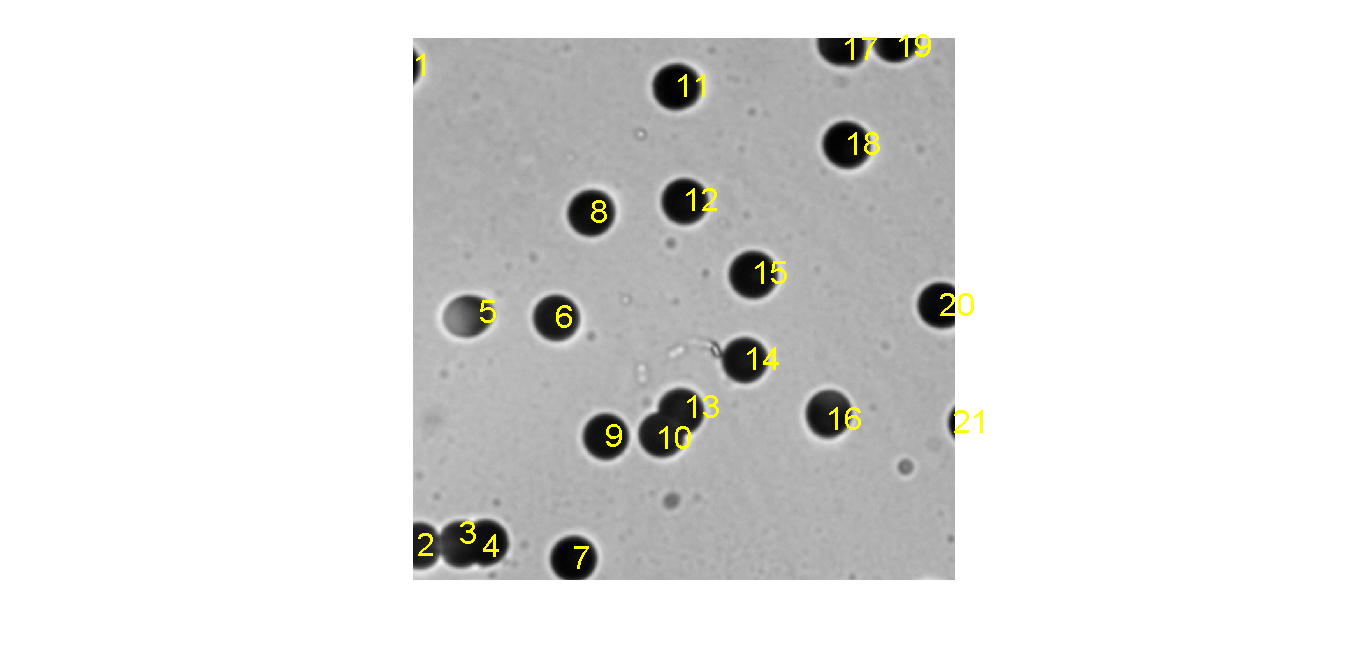}
\caption{(a) 6 h etched $~^{252}$Cf exposed CR-39. Here track density is $\sim10^9$ m$^{-2}$.(b) Identification and counting of etch-pit openings at the overlapped region as well as at the extreme edges of the figure.}
\label{fig7} 
\end{figure}

We have tested our proposed method with $~^{32}$S ion exposed PET for both normal (resulting circular etch-pit opening (Fig.\ref{fig4}(a))) and angular (resulting elliptical etch-pit opening(Fig.\ref{fig5}(a))) incidences. Besides, in order to check the robustness of the proposed method, we have applied it on the open-air exposed CR-39 having degraded surface quality and containing etch-pits openings of different shapes and sizes (Fig.\ref{fig6}(a)) and $~^{252}$Cf exposed CR-39 having higher track density and overlap region (Fig.\ref{fig7}(a)). In each of the cases, the surfaces are riddled with different types of defects, but as shown in Fig.\ref{fig4}(b), Fig.\ref{fig5}(b), Fig.\ref{fig6}(b) and Fig.\ref{fig7}(b), the same algorithm yields good results and most of the tracks are correctly recognized.

\par
\begin{figure}[h]
\centering
\includegraphics[width=400px,height=200px]{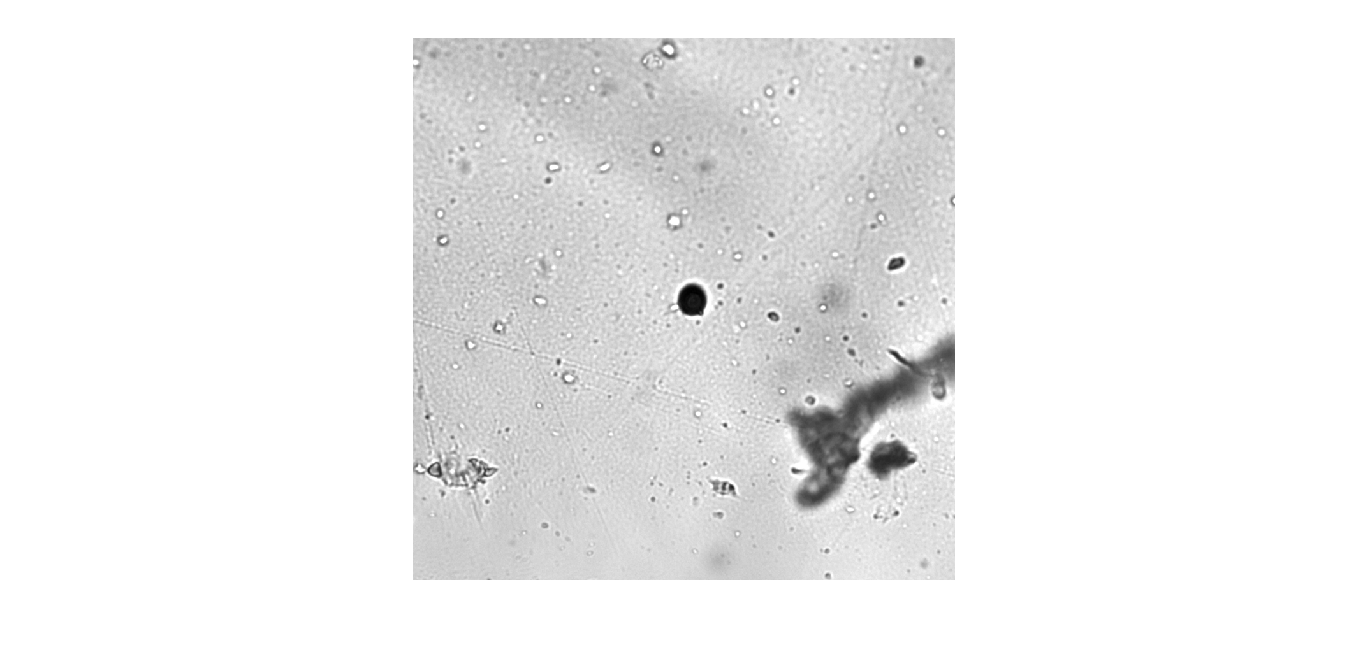}
\includegraphics[width=400px,height=200px]{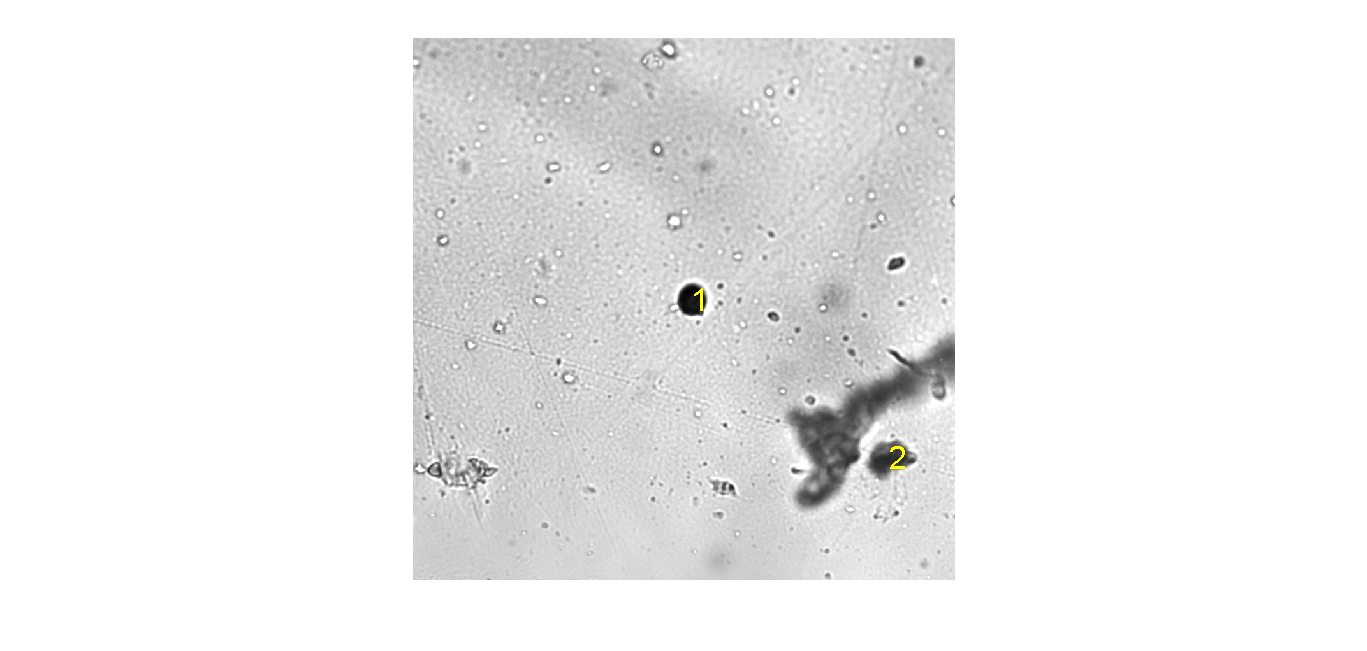}
\caption{(a) One circular etch-pit opening due to normally incident $~^{32}$S exposed PET at the middle of the figure and a defect of similar size and shape at the bottom-right side. (b) One typical case among the very few misjudgments where a defect is wrongly counted.}
\label{fig8} 
\end{figure}

We have examined hundreds of similar samples and in most of the cases, approximately all the etch-pits are identified, making error percentage extremely small as given in Table 1. It may be mentioned here that the proposed method is based on the convolution with a suitable mask. Judicial choice of the mask size plays the key role here for identifying the etch-pit openings. In the cases where the size and/or shape of defects of the material mimic the actual etch-pit opening (as shown in bottom-right of Fig.\ref{fig8}(a)), those artifacts may be wrongly counted (Fig.\ref{fig8}(b)). 

\begin{table*}
\centering 
{
\begin{tabular}{|c|c|c|c|} \hline
\bf Images of etch-pit       & \bf Manual   & \bf Automated  & \bf Percentage \\
\bf  openings from   & \bf count    & \bf count      & \bf   error             \\ \hline

Accelerator exposure        & $301$          & $302$             & $0.33\%$           \\ \hline
Open-air exposure        & $152$         & $147$             & $3.3\%$            \\ \hline
\end{tabular}}
{
\caption{Results of the manual and automatic counts of etch-pits on 58 and 53 images of accelerator exposed PET and open-air exposed CR-39 films respectively.}}
\label{table1}
\end{table*}

\section{Comparison with other methods}
Any shape detection algorithm such as Hough transform or shape fitting etc. works for a definite shape. Even if there are slight deviations from the actual shape, most of these algorithms fail to detect the shape. In the present case, etch-pit openings can be both circular and elliptical but with wide variations in their size. Our attempts with conventional image analysis methods didn't yield any good results. These techniques can't be easily applied for counting without using morphological techniques as well. Contrary to conventional methods, using the proposed convolution method, the detection of a peak at the location of the etch-pits is much better and it simultaneously provides its coordinates as well. The proposed method is simpler yet more robust than some well-known shape detecting algorithms like Hough Transform, morphological operations, watershed segmentation~\citep{1397242,Sharif_2016,774169} etc. A few other advantages are listed below:

\begin{itemize}
  \item No simultaneous frequency domain processing is required for this technique.
  \item Even at the overlapping regions of two or more etch-pit openings, which are difficult to handle with other image analysis methods, tracks can be easily recognized and counted by this technique.
  \item By virtue of this technique, it is possible to identify etch-pit openings even at the extreme edges of the image frame.
\end{itemize}

\section{Conclusion}
A novel image analysis technique based on convolution is shown to generate much better results compared to many other techniques for etch-pit detection in NTDs. It promises to substantially speed up the task of track identification and analysis, which is the chief technical challenge for any experiments employing large area NTD arrays. To the best of our knowledge, this is the first application of this particular technique to the problem of NTD image analysis.

 \section*{Acknowledgments}
 
The authors are grateful to Dr. Debapriyo Syam for many useful suggestions. 
KP and JC wish to acknowledge TEQIP-III, University of Calcutta and Department of Applied Optics and Photonics, University of Calcutta. 
RB wishes to acknowledge Project No. SERB/PHY/2016/041 and SB/S2/RJN-29/2013 for financial support. This work is partially funded by IRHPA (Intensification of Research in
High Priority Areas) Project (IR/S2/PF-01/2011 dated 26.06.2012)
of the Science and Engineering Research Council (SERC), DST,
Government of India, New Delhi.

\section*{References}
\bibliography{NTDimage}

\end{document}